\documentclass[review]{elsarticle}

\usepackage{graphicx}
\usepackage{latexsym, amsmath, verbatim, graphics, epsfig, amssymb, color, setspace, url}
\usepackage{setspace}
\usepackage{subfig}
\newtheorem{theorem}{Theorem}[section]

\begin{document}
	
	\begin{frontmatter}
		
		\title{A SEIRUC mathematical model for transmission dynamics of COVID-19}
		
		\author{P. Tamilalagan$^{\ast,a}$, B. Krithika$^a$, P. Manivannan$^b$}
		\address{$^a$Department of Mathematics, School of Engineering, Amrita Vishwa Vidyapeetham,\\
			Coimbatore, Tamil Nadu, India.}
		\address{$^c$ Department of Mathematics, Mepco Schlenk Engineering College, Sivakasi, \\
			Virudhunagar, Tamil Nadu-626005, India.}
		\cortext[mycorrespondingauthor]{Corresponding author, Email: p$\_$tamilalagan@cb.amrita.edu, mathematicstamil@gmail.com}
		
		
		\begin{abstract}
			The world is still fighting against COVID-19, which has been lasting for more than a year. Till date, it has been a greatest challenge to human beings in fighting against COVID-19 since, the pathogen SARS-COV-2 that causes COVID-19 has significant biological and transmission characteristics when compared to SARS-COV and MERS-COV pathogens. In spite of many control strategies that are implemented to reduce the disease spread, there is a rise in the number of infected cases around the world. Hence, a mathematical model which can describe the real nature and impact of COVID-19 is necessary for the better understanding of disease transmission dynamics of COVID-19. This article proposes a new compartmental SEIRUC mathematical model, which includes the new state called convalesce (C). The basic reproduction number $\mathcal{R}_0$ is identified for the proposed model. The stability analysis are performed for the disease free equilibrium ($\mathcal{E}_0$) as well for the endemic equilibrium ($\mathcal{E}_*$) by using the Routh-Hurwitz criterion. The graphical illustrations of the proposed mathematical results are provided to validate the theoretical results.
		\end{abstract}
		
		\begin{keyword}
			\texttt{COVID-19}\sep Mathematical model \sep Stability analysis \sep Dynamical systems.
			\MSC[2010] 34K20 \sep 37N25 \sep 34A34 \sep 70K20
		\end{keyword}
		
	\end{frontmatter}
	
	
	\section{Introduction}
	
	%
	%
	Mathematical modelling of infectious disease using nonlinear dynamical systems can give significant insight into the transmission dynamics or dynamical behaviour of disease spread. Epidemiological modelling of diseases has improved drastically over the past decades and continues to rise up in several fields \cite{FTK3RF1}-\cite{FTK3RF3}. In particular, the differential equation models have been utilized to develop biological and physical problems \cite{FTK3RF4}-\cite{FTK3RF9}. COVID-19 is a disease caused by a new virus, which is generating a pandemic worldwide and needs a model taking into account its known specification characteristics. Due to the significance and powerful nature of differential equation models in epidemiology, some recent studies in literature have considered the mathematical modelling of the COVID-19 pandemic  using nonlinear differential equations \cite{FTK3RF10}-\cite{FTK3RF15}. Since pandemics are large-scale outbreaks of infectious disease, it can produce an important risk to human life over a wide geographic area and can cause economic and social disruption. Mathematical models comprising derivatives aid in estimating the effect of precautionary measures adopted against novel coronavirus. Recently, few mathematical models have been investigated by many researchers to understand the transmission dynamics of COVID-19 pandemic and some of these are listed in our references. There are some mathematical models in the literature that try to describe the dynamics of the evolution of COVID-19 \cite{FTK3RF10}-\cite{FTK3RF15}. Other works \cite{FTK3RF11},\cite{FTK3RF15} propose SIR and SEIR type models with little variations. Piu Samui et al. \cite{FTK3RF10} proposed a SAIU model for the spread of COVID-19 using data from case study of India, taking into account the asymptomatic, reported symptomatic infectious and unreported symptomatic infectious class. Abdullah et al.\cite{FTK3RF12} introduced a mathematical model by including resistive class together with quarantine class and use it to investigate the transmission dynamics of novel corona virus disease. Liu et al.\cite{FTK3RF14} introduced a COVID-19 epidemic model taking into account the latency period. As identified by the World Health Organisation (WHO), the mathematical models, mainly those formulated on a timely basis plays a vital role in allowing public health decision and decision makers with evidence-based statistics \cite{FTK3RF16}-\cite{FTK3RF18}.\\
	\indent According to Worldometer data, 159,068,471 total confirmed cases, 3,308,750 confirmed deaths and 136,633,409 recovered cases has been recorded throughout the world as of May 10, 2021 \cite{FTK3RF19}. Also, as per the report of World Health Organization (WHO) as of 23rd December, 2020, the individuals infected to SARS-Cov-2 virus, which causes COVID-19, develop antibodies after infection \cite{FTK3RF16}. It has been reported that, the infected individuals who have even severe and mild disease also develop these antibodies. Hence, Serological studies and research are underway to recognize the stability of this immune response and also to investigate, how far these antibodies last.\cite{FTK3RF18}\\
	\indent This present article try to incorporate the antibody response of immune system to the COVID-19 disease as a separate compartment namely, Convalesce. Specifically, the immune system of the individuals belonging to the Convalesce class is strong enough to be safe from pandemic, so that they are not infected again. It also accounts the population recovered by taking the home herbal medicine or region-specific traditional medicines in various forms in different countries. Traditional Chinese medicine substances used in clinical trials includes Polygonum cuspidatum(also known as Asian knotweed), Honey suckle, Ligustrum lucidum(an evergreen tree) etc,\cite{FTK3RF20}. The World Health Organisation has also recommended inclusion of traditional medicine in its COVID-19 strategic preparedness and response \cite{FTK3RF21}. Indian traditional system, Ayurveda has a clear concept of the cause and treatment of pandemics, \cite{FTK3RF22} provides information on the potential antiviral traditional medicines along with their immunomodulatory pathways and also described seven most important Indian traditional plants with antiviral properties. Recently, an antiviral drug, Clevira has been approved by Government of India, as a supporting measure for mild to moderate condition of Covid-19. The trial outcomes revealed that Clevira has shown 86 per cent recovery rate on fifty days of treatment in mild to moderate Covid-19 cases \cite{FTK3RF23}. Based on these discussions, the individuals belonging to the Convalesce class are assumed to be immune to reinfection. \\
	\indent The coming sections of this article is sorted out as follows: We propose a dynamical model and the model description in section 2. The qualitative properties of the proposed SEIRUC model have been discussed in section 3. In the same section, the basic reproduction number for the SEIRUC model have been computed and the stability analysis of both disease free equilibrium and endemic equilibrium points have been performed. In section 4, we conduct sensitivity analysis for the basic reproduction number. In section 5, we present numerical simulation to verify our analytical findings and a discussion in section 6 concludes our manuscript.
	\section{Model Description}
	\noindent The SEIRUC mathematical model for transmissions dynamics of COVID-19, which is under consideration, is given below,
	\begin{eqnarray}\label{e1}
		\frac{dS(t)}{dt}&=&\Lambda-\frac{\beta S(t)}{N(t)}(I(t)+U(t)+R(t))-\delta S(t),\nonumber\\
		\frac{dE(t)}{dt}&=&\frac{\beta S(t)}{N(t)}(I(t)+U(t)+R(t))-(a+\delta_1+c_1)E(t),\nonumber\\
		\frac{dI(t)}{dt}&=&a E(t)-(\gamma+\delta_2+c_2)I(t),\\
		\frac{dR(t)}{dt}&=& \gamma q I(t)-(\delta_3+c_3)R(t)+\eta U(t),\nonumber\\
		\frac{dU(t)}{dt}&=& \gamma(1-q) I(t)-(\delta_4+c_4)U(t)-\eta U(t),\nonumber\\
		\frac{dC(t)}{dt}&=& c_1 E(t)+c_2 I(t)+c_3 R(t)+c_4 U(t)-\delta_5 C(t) \nonumber
	\end{eqnarray}
	where $N(t)=S(t)+E(t)+I(t)+R(t)+C(t)$ is the total size of the population, $S(t)$ denotes the population of individuals susceptible to the infection, (i.e) who are uninfected but vulnerable to the infection at any time $t$. $E(t)$ denotes the population of asymptomatic noninfectious individuals, (i.e) who are infected but not having any symptoms and not capable of transmitting the infection to others at any time $t$. $I(t)$ denotes the population of asymptomatic but infectious individuals, (i.e) who are infected and not having any symptoms but capable of transmitting the infection to others at any time $t$. $R(t)$ denotes the population of reported symptomatic infectious individuals, (i.e) who are infected and having symptoms of the infection also capable of transmitting the infection to others at any time $t$ moreover they are reported to medical authorities and undergone treatment through hospitals. $U(t)$ denotes the population of unreported symptomatic infectious individuals, (i.e) who are infected and having symptoms of the infection also capable of transmitting the infection to others at any time $t$ but they did not reported about their infection to any medical authorities and not undergone treatment through hospitals. $C(t)$ denotes the population of convalesce individuals, who are recovered from the infection and not susceptible to the infection again. Here $t\geq t_0$ is time in days further $t_0$ represents the starting date of the endemic. It must be emphasized that some parameters are dependent of time but the theoretical results of the model (\ref{e1}) are performed with constant parameters.
	
	In the above model (\ref{e1}), $\Lambda$ denotes net inflow of susceptible or uninfected population, the uninfected population $S(t)$ becomes infected at a rate $\beta$ by exposed with infected infectious individuals namely $I, \ R,$ and $U$, the natural decay rate of $S(t)$ is $\delta$. In the population $E(t)$, the parameter $a$ denotes the rate at which the infected noninfectious population $E(t)$ becomes infectious $I(t)$. The asymptomatic infectious population $I(t)$ is asymptomatic for the period $\frac{1}{\gamma}$. The populations $E(t)$ and $I(t)$ are recovered respectively at a rate $c_1$ and $c_2$ due to protective immune response to the infection by human immune system and as the infection is not acute at this stage, it also depends on the age of the individual. The natural decay rates of the populations $E(t)$ and $I(t)$ are respectively $\delta_1$ and $\delta_2$, this accounts a fraction of death's caused by COVID-19 without any symptoms. The asymptomatic infectious population $I(t)$ becomes reported symptomatic infectious $R(t)$ at a rate $\gamma q$, where $q$ is the fraction of asymptomatic infectious that become reported symptomatic infectious, correspondingly $(1-q)\gamma$ is the rate of asymptomatic infectious population $I(t)$ becomes unreported symptomatic population $U(t)$. $\eta$ is the rate at which the unreported symptomatic infectious population, $U(t)$ becomes reported symptomatic infectious population, $R(t)$. The recovery rate of the population $R(t)$ is $c_3$, it accounts, the recovered population due to severe treatment by hospitals. The natural decay rates of the populations $R(t)$ and $U(t)$ are respectively $\delta_3$ and $\delta_4$. The population $U(t)$ recovered at a rate $c_4$. The mortality rate of the population $C(t)$ is $\delta_5$, it accounts the individuals recovered from COVID-19 but died because of other reasons/diseases.
	\section{Qualitative Properties of the SEIRUC Model}
For the SEIRUC model to be epidemiologically realistic, it is necessary to prove that the model solutions are positive under non-negative initial conditions.
\subsection{Positivity of solutions}
\begin{theorem}
	Let $S_0\geq0, E_0\geq0, I_0\geq0, R_0\geq0, U_0\geq0, C_0\geq0$. The solution of system with $(S(0),E(0),I(0),R(0),U(0),C(0))=(S_0, E_0, I_0, R_0, U_0, C_0)$ is non-negative ie., $S(t)\geq0$, $E(t)\geq0$, $I(t)\geq0$, $R(t)\geq0$, $U(t)\geq0$, $C(t)\geq0$ for $t>0$.
\end{theorem}
\textbf{Proof.} Let $x(t)=(S(t),E(t),I(t),R(t),U(t),C(t))$ be the solution of system (\ref{e1}) under initial conditions $x_0=(S_0, E_0, I_0, R_0, U_0, C_0)\geq 0$.\\
Consider
\begin{align*}
\frac{dS(t)}{dt}=\Lambda-\frac{\beta S(t)}{N}(I(t)+U(t)+R(t))-\delta S(t).
\end{align*}
\begin{align*}
\frac{dS(t)}{dt}=\Lambda-\Upsilon_1(t)S(t).
\end{align*}
where\indent $\Upsilon_1(t)=\frac{\beta}{N}(I(t)+U(t)+R(t))+\delta$.\\
Thereafter, we obtain the following expression.
\begin{align*}
S(t)=S_0e^{-\int_{0}^{t}\Upsilon_1(s)ds}+\Lambda e^{-\int_{0}^{t}\Upsilon_1(s)ds}\int_{0}^{t}e^{\int_{0}^{s}\Upsilon_1(u)du}ds>0.	
\end{align*}
This implies, $S(t)$ is non-negative for all t. \\
Similarly, it can also be shown that $E(t), I(t), R(t), U(t), C(t) > 0$ for all $t>0.$, which implies that the disease is uniformly persistent for every positive solution.
\begin{theorem}
Let $(S(t),E(t),I(t),R(t),U(t),C(t))$ be the solution of the system with initial conditions $(S_0,E_0,I_0,R_0,U_0,C_0)$ and let $\mu=$min$(\delta,\delta_1,\delta_2,\delta_3,\delta_4,\delta_5)$. \\
The compact set, $\Phi=\left\lbrace(S(t),E(t),I(t),R(t),U(t),C(t))\in\mathbb{R}^6_+, N\leq\frac{\Lambda}{\mu}\right\rbrace$ is positively invariant set and attracts all solutions in $\mathbb{R}^6_+.$
\end{theorem}
\textbf{Proof.} Here, $N(t)=S(t)+E(t)+I(t)+R(t)+U(t)+C(t).$ Then,
\begin{align*}
\frac{dN}{dt}&\leq \Lambda-\mbox{min}(\delta,\delta_1\delta_2,\delta_3,\delta_4,\delta_5)N(t)\\
&=\Lambda-\mu N(t)\\
\Longrightarrow\hspace{1cm}\frac{dN}{dt}+\mu N&\leq\Lambda.
\end{align*}
Solving the above equation, we obtain $0<N\leq\frac{\Lambda}{\mu}+(N(0)-\frac{\Lambda}{\mu})e^{-\mu t}$.\\
Hence, $\lim_{t\rightarrow\infty} Sup N(t)\leq\frac{\Lambda}{\mu}$.\\
Thus, $\Phi$ is positively invariant and attractive set. Therefore, all the feasible soluions in the model converge in $\Phi$.
\subsection{Equilibrium Points}
	The proposed model has two exclusively different steady states namely endemic free or disease free equilibrium ${\cal E}_0=\left(S_0,0,0,0,0,0\right)$ and the endemic equilibrium ${\cal E}_\ast=(S_\ast,E_\ast,I_\ast,R_\ast,U_\ast,C_\ast)$, where
	\sep
	\begin{align*}
		E_{\ast}&=\frac{\Lambda-\delta S_{\ast}}{a+\delta_1+c_1}, \quad I_{\ast}=\frac{aE_{\ast}}{\gamma+\delta_2+c_2}, \quad R_{\ast}=\frac{\gamma q I_{\ast}+\eta U_{\ast}}{\delta_3+c_3},  \\
		U_{\ast}&=\frac{\gamma(1-q)I_{\ast}}{\eta+\delta_4+c_4},\quad C_{\ast}=\frac{c_1E_{\ast}+c_2I_{\ast}+c_3R_{\ast}+c_4U_{\ast}}{\delta_5},
	\end{align*}
	and $S_{\ast}$ is given by the quadratic equation
	\begin{eqnarray}\label{e2}
	AS^2_{\ast}+BS_{\ast}+\Lambda=0
	\end{eqnarray}
	here
	\begin{align*}
		A&=\frac{a\delta\beta[(\delta_3+c_3)(\eta+\delta_4+c_4)+\gamma(1-q)(\delta_3+c_3)+\gamma q(\eta+\delta_4+c_4)+\gamma(1-q)\eta]}{N(a+\delta_1+c_1)(\gamma+\delta_2+c_2)(\delta_3+c_3)(\eta+\delta_4+c_4)},\\
		B&=-\frac{a\Lambda\beta[(\delta_3+c_3)(\eta+\delta_4+c_4)+\gamma(1-q)(\delta_3+c_3)+\gamma q(\eta+\delta_4+c_4)+\gamma(1-q)\eta]}{N(a+\delta_1+c_1)(\gamma+\delta_2+c_2)(\delta_3+c_3)(\eta+\delta_4+c_4)}-\delta
	\end{align*}
	The roots of (\ref{e2}) are $S_{\ast}=\frac{\delta}{A}$ and $S_{0}=\frac{\Lambda}{\delta}$. Further, the second root $S_{0}=\frac{\Lambda}{\delta}$ entails the endemic free steady state ${\cal E}_0$.
	\subsection{Stability Analysis of Endemic free Steady state ${\cal E}_0$}
	\begin{theorem}
		The dynamical system (\ref{e1}) is locally asymptotically stable at the endemic-free equilibrium, $\mathcal{E}_0=(\frac{\Lambda}{\delta},0,0,0,0,0)$ when $\mathcal{R}_0<1$ and unstable if $\mathcal{R}_0>1.$
	\end{theorem}
	\textit{3.3.1 Basic Reproduction Number $\mathcal{R}_0$}\\
	\indent The basic reproduction number $\mathcal{R}_0$ is used to estimate the average number of new infections produced by infective individual over the period of infection. Every epidemiological models have basic reproduction number $\mathcal{R}_0$ as the threshold parameter for the local stability of the model. It is reported in the literature that, if $\mathcal{R}_0<1$ then the disease free equilibrium is locally asymptotically stable and the disease can not spread in large numbers, since over the entire course of infection one infected individual produces less than one new infected individual and the disease cannot persist. In contrast, if $\mathcal{R}_0>1$ then on average more than one new infection could be caused by infected individual and the disease may invade the population.
	
	The basic reproduction number $\mathcal{R}_0$ corresponding to the proposed SEIRUC model (\ref{e1}) is obtained by using the procedure in \cite{FTK3RF23} and \cite{FTK3RF24}. It is given by the spectral radius of the next generation matrix $FV^{-1}$ in symbolic representation $$\mathcal{R}_0=\rho(FV^{-1})$$
Here, we consider the matrices representing the production of new infection and transition part of our proposed SEIRUC model as follows\\
\begin{align*}
\mathcal{F}=\begin{bmatrix}
\frac{\beta S(t)}{N}(I(t)+U(t)+R(t))\\
0\\
0\\
0
\end{bmatrix}, \mathcal{V}=\begin{bmatrix}
(a+\delta_1+c_1)E(t)\\
-aE(t)+(\gamma+\delta_2+c_2)I(t)\\
-\gamma qI(t)+(\delta_3+c_3)R(t)-\eta U(t)\\
-\gamma(1-q)I(t)+(\delta_4+c_4+\eta)U(t)
\end{bmatrix}
\end{align*}From which we obtain
	\begin{align*}
		F&=\left(
		\begin{array}{cccc}
			0 & \frac{\beta S_0}{N} & \frac{\beta S_0}{N} & \frac{\beta S_0}{N} \\
			0 & 0 & 0 & 0 \\
			0 & 0 & 0 & 0 \\
			0 & 0 & 0 & 0 \\
		\end{array}
		\right),
		\quad V&=\left(
		\begin{array}{cccc}
			a+\delta_1+c_1 & 0 & 0 & 0 \\
			-a & \gamma+\delta_2+c_2 & 0 & 0 \\
			0 & -q\gamma & \delta_3+c_3 & -\eta \\
			0 & -(1-q)\gamma & 0 & \delta_4+c_4+\eta \\
		\end{array}
		\right)
	\end{align*}
	where, the matrices $F$ and $V$ are the Jacobians of the matrices $\mathcal{F}$ and $\mathcal{V}$ respectively. Thus, we have,
	$$\mathcal{R}_0=\frac{a\beta\Lambda\mathcal{P}}{N\delta\theta_1\theta_2\theta_3\theta_4}.$$\\ Where, \quad $\mathcal{P}=\gamma(1-q)(\eta+\theta_3)+\theta_4(\theta_3+q\gamma),\quad \theta_1=a+\delta_1+c_1,\quad \theta_2=\gamma+\delta_2+c_2,\quad\theta_3=\delta_3+c_3$\\
	$\theta_4=\delta_4+c_4+\eta.$
	\begin{figure}
		\begin{minipage}[b]{.4\textwidth}
			\centering
			\includegraphics[width=1\textwidth]{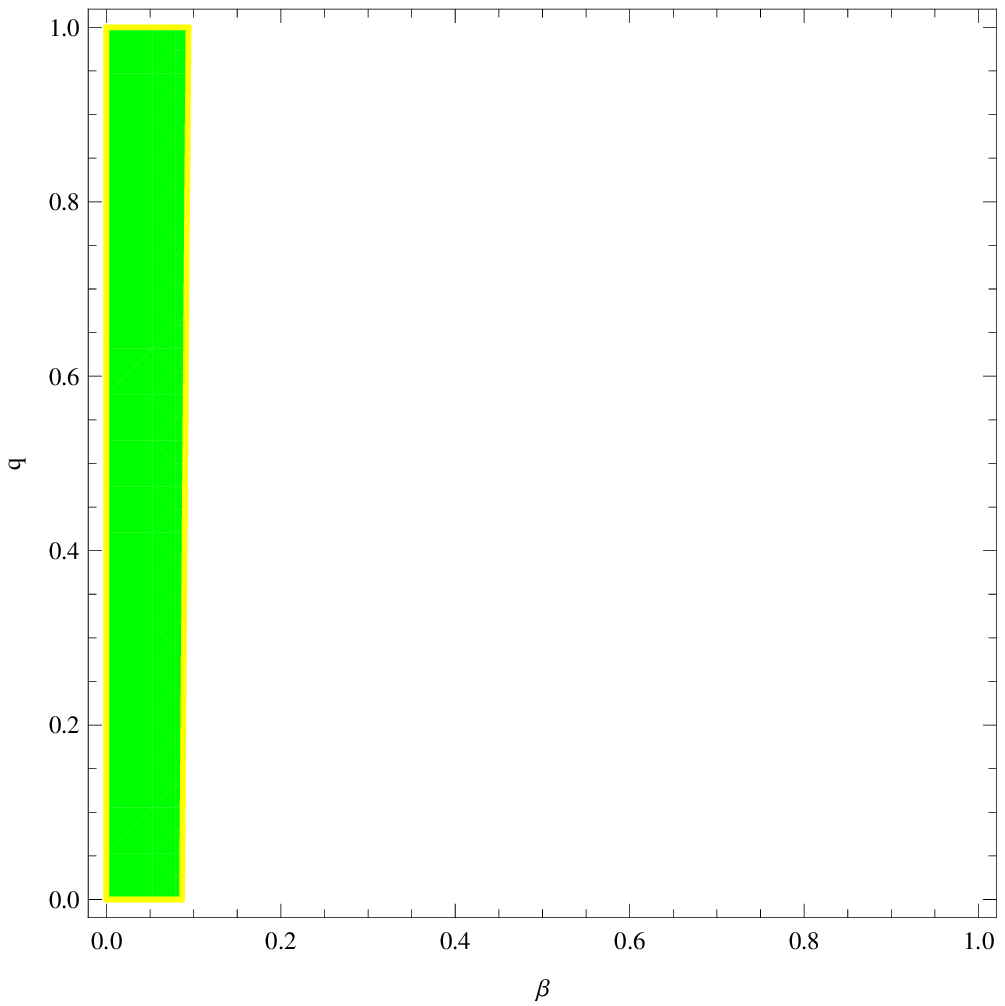}
			\caption{Region plot for $\mathcal{R}_0<1$ with respect to parameters $\beta$ and $q$.}
			\label{Fig.Figure2}
		\end{minipage}
		\hfill
		\begin{minipage}[b]{.4\textwidth}
			\centering
			\includegraphics[width=1\textwidth]{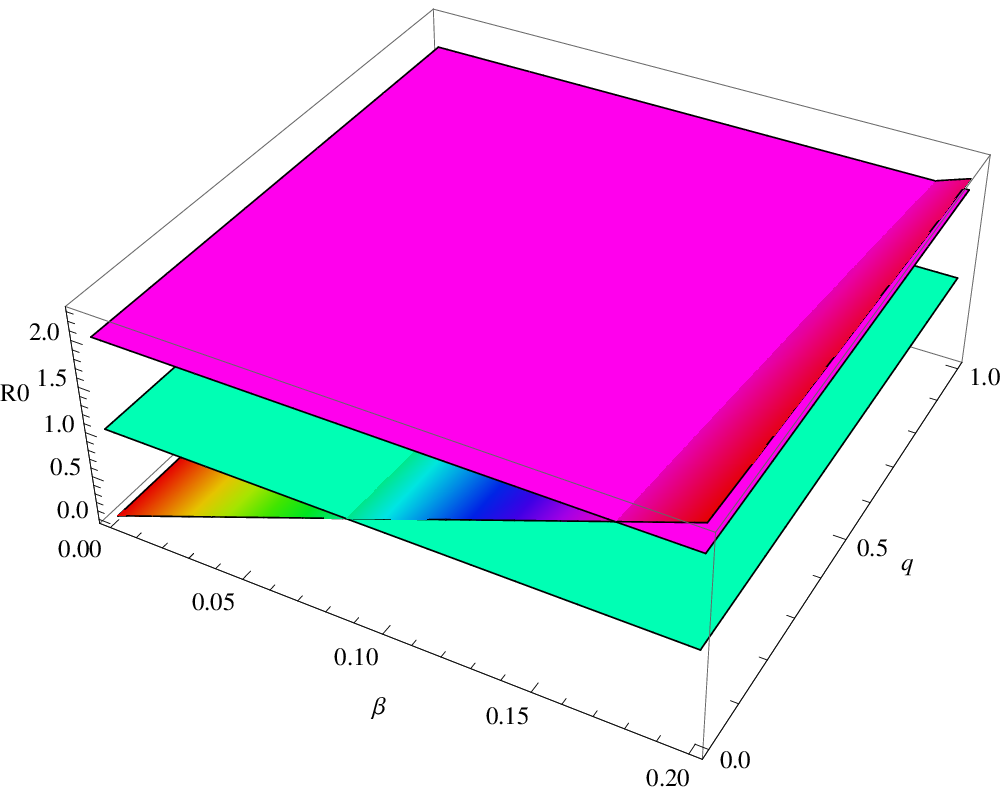}
			\caption{Plot of basic reproduction number, $\mathcal{R}_0$ with respect to parameters $\beta$ and $q$.}
			\label{Fig.Figure3}
		\end{minipage}
	\end{figure}
\subsection{Stability Analysis of Endemic Steady state ${\mathcal E}_\ast$}
In this subsection, the existence and stability conditions for the endemic steady state are presented. The endemic steady state can be rewritten as follows
\begin{align*}
S_*&=\frac{\Lambda}{\delta\mathcal{R}_0}\\
E_*&=\frac{(\mathcal{R}_0-1)N\delta\theta_2\theta_3\theta_4}{\mathcal{P}a\beta}\\
I_*&=\frac{(\mathcal{R}_0-1)N\delta\theta_3\theta_4}{\mathcal{P}\beta}\\
R_*&=\frac{(\mathcal{R}_0-1)N\delta(\eta\gamma(1-q)+q\theta_4\gamma)}{\mathcal{P}\beta}\\
U_*&=\frac{\gamma(1-q)(\mathcal{R}_0-1)N\delta\theta_3}{\mathcal{P}\beta}\\
C_*&=\frac{(\mathcal{R}_0-1)N\delta(a\gamma(1-q)(c_3\eta+c_4\theta_3)+\theta_3\theta_4(ac_2+c_1\theta_2)+ac_3q\theta_4\gamma)}{\mathcal{P}a\beta\delta_5}
\end{align*}
The term, $\mathcal{P}=\gamma(1-q)(\eta+\theta_3)+\theta_4(\theta_3+\gamma q)$ is positive since $q$ is the fraction of asymptomatic infectious that become reported symptomatic infectious, it lies between 0 and 1. Hence, it is obvious from the above that, the positive endemic steady state of system (\ref{e1}) exist only when $\mathcal{R}_0>1$. Hence, the model system $(\ref{e1})$ has an endemic equilibrium point whenever $\mathcal{R}_0>1$ and has no endemic steady state for $\mathcal{R}_0\leq1.$ The stability results are proved using Routh-Hurwitz criterion in the following theorem.
\begin{theorem} The dynamical system $(\ref*{e1})$ is locally asymptotically stable at the endemic equilibrium point $\mathcal{E}_*=(S_*,E_*,I_*,R_*,U_*,C_*)$, for $\mathcal{R}_0>1$ and further if the following inequalities hold
\end{theorem}
\begin{itemize}
	\item[$A_1:$] $\Gamma_1>\Gamma_2$
	\item[$A_2:$]$\Gamma_3>\Gamma_4$
	\item[$A_3:$]$\mathcal{R}_0\delta b_3>\Gamma_5$
\end{itemize}
$where$
\begin{align*}
\Gamma_1&=(\mathcal{R}_0\delta+b_1)(\mathcal{R}_0\delta b_1+b_2)+\frac{b_4}{\mathcal{P}}(\delta+\gamma+\theta_3+\theta_4)
\end{align*}
\begin{align*}
\Gamma_2&=(\mathcal{R}_0\delta+b_1)\frac{b_4}{\mathcal{P}}+\mathcal{R}_0\delta b_2+b_3\\
\Gamma_3&=\left(\frac{\Gamma_1-\Gamma_2}{\mathcal{R}_0\delta+b_1}\right)(\mathcal{R}_0\delta b_2+b_3)+(\mathcal{R}_0\delta+b_1)\frac{b_4\delta}{\mathcal{P}}(\gamma+\theta_3+\theta_4)+\delta b_4(\mathcal{R}_0-1)\\
\Gamma_4&=\left(\frac{\Gamma_1-\Gamma_2}{\mathcal{R}_0\delta+b_1}\right)\frac{b_4}{\mathcal{P}}(\delta+\gamma+\theta_3+\theta_4)+(\mathcal{R}_0\delta+b_1)(\mathcal{R}_0\delta b_3)\\
\Gamma_5&=\frac{b_4\delta}{\mathcal{P}}(\gamma+\theta_3+\theta_4)+\frac{\delta b_4(\mathcal{R}_0-1)}{\mathcal{R}_0\delta+b_1}\left(1+\frac{(\Gamma_1-\Gamma_2)^2}{(\Gamma_3-\Gamma_4)(\mathcal{R}_0\delta+b_1)}\right).	
\end{align*}
\textbf{Proof.}\indent The stability nature of the endemic steady state $\mathcal{E}^*=(S^*,E^*,I^*,R^*,U^*,C^*)$ can be determined using the eigenvalues of the Jacobian matrix $J(\mathcal{E}^*)$. Hence, evaluating the Jacobian matrix around the endemic equilibrium $\mathcal{E}_*(S_*,E_*,I_*,R_*,U_*,C_*)$, we obtain,
\begin{eqnarray}
J(\mathcal{E}_*)=\left(\begin{matrix}
-\mathcal{R}_0\delta&0&-\frac{b_4}{a\mathcal{P}}&-\frac{b_4}{a\mathcal{P}}&-\frac{b_4}{a\mathcal{P}}&0\\
(\mathcal{R}_0-1)\delta&-\theta_1&\frac{b_4}{a\mathcal{P}}&\frac{b_4}{a\mathcal{P}}&\frac{b_4}{a\mathcal{P}}&0\\
0&a&-\theta_2&0&0&0\\
0&0&\gamma q&-\theta_3&\eta&0\\
0&0&\gamma(1-q)&0&-\theta_4&0\\
0&c_1&c_2&c_3&c_4&-\delta_5
\end{matrix}\right)
\end{eqnarray}
Taking
\begin{align*}
	b_1&=\theta_1+\theta_2+\theta_3+\theta_4\\
	b_2&=\theta_1\theta_2+\theta_1\theta_3+\theta_1\theta_4+\theta_2\theta_3+\theta_2\theta_4+\theta_3\theta_4\\
	b_3&=\theta_1\theta_2\theta_3+\theta_1\theta_2\theta_4+\theta_2\theta_3\theta_4+\theta_1\theta_3\theta_4\\
	b_4&=\theta_1\theta_2\theta_3\theta_4.
\end{align*}
From the characteristic equation of the Jacobian matrix $J(\mathcal{E}_*)$, we obtain, one of the eigenvalues as $-\delta_5$ and the remaining eigenvalues are the roots of the following polynomial
\begin{align*}
x^5+\widehat{a}_1x^4+\widehat{a}_2x^3+\widehat{a}_3x^2+\widehat{a}_4x+\widehat{a}_5=0.
\end{align*}
Where
\begin{align*}
\widehat{a}_1&=\mathcal{R}_0\delta+b_1\\
\widehat{a}_2&=\mathcal{R}_0\delta b_1+b_2-\frac{b_4}{\mathcal{P}}\\
\widehat{a}_3&=\mathcal{R}_0\delta b_2+b_3-\frac{b_4}{\mathcal{P}}(\delta+\gamma+\theta_3+\theta_4)\\
\widehat{a}_4&=\mathcal{R}_0\delta b_3-\frac{b_4\delta}{\mathcal{P}}(\gamma+\theta_3+\theta_4)\\
\widehat{a}_5&=\delta b_4(\mathcal{R}_0-1).
\end{align*}
Let $\widehat{s}_1=\widehat{a}_2-\frac{\widehat{a}_3}{\widehat{a}_1}$,\hspace{0.2cm}$\widehat{s}_2=\widehat{a}_4-\frac{\widehat{a}_5}{\widehat{a}_1}$,\hspace{0.2cm} $\widehat{t}_1=\widehat{a}_3-\frac{\widehat{a}_1\widehat{s}_2}{\widehat{s}_1}$,\hspace{0.2cm} $\widehat{t}_2=\widehat{a}_5$,\hspace{0.2cm} $\widehat{d}_1=\widehat{s}_2-\frac{\widehat{s}_1\widehat{t}_2}{\widehat{t}_1}$.
Then, according to Routh-Hurwitz criterion, when $\mathcal{R}_0>1$, the system $(\ref{e1})$ has eigenvalues with negative real parts if $\widehat{a}_1>0$, $\widehat{s}_1>0$, $\widehat{t}_1>0$, $\widehat{d}_1>0$ and $\widehat{a}_5>0.$ \\
Here, we obtain$$
\widehat{s}_1=\frac{\Gamma_1-\Gamma_2}{\mathcal{R}_0\delta+b_1}>0\hspace{0.3cm}\mbox{if and only if}\hspace{0.2cm} A_1\hspace{0.2cm}\mbox{holds}$$\\
Similarly, we obtain $$\widehat{t}_1=\frac{\mathcal{R}_0\delta+b_1}{\Gamma_1-\Gamma_2}(\Gamma_3-\Gamma_4)\hspace{0.2cm}\mbox{and}\hspace{0.2cm}\widehat{d}_1=\mathcal{R}_0\delta b_3-\Gamma_5.$$
Hence, it can be seen that $\widehat{t}_1>0$ and $\widehat{d}_1>0$ whenever $A_2$ and $A_3$ holds and the last coefficient $\widehat{a}_5>0$ whenever $\mathcal{R}_0>1$. Thus, by Routh-Hurwitz criterion, when $\mathcal{R}_0>1$, the endemic steady state, $\mathcal{E}_*=(S_*,E_*,I_*,R_*,U_*,C_*)$ is locally asymptotically stable if $\widehat{s}_1>0$, $\widehat{t}_1>0$ and $\widehat{d}_1>0$ which holds whenever $A_1$, $A_2$ and $A_3$ holds. \hfill$\square$.
\section{Sensitivity Analysis}
In literature, sensitivity analysis is proposed to understand the relative importance of the different factors responsible for transmission and prevalence of the disease. In order to reduce the disease transmission, it is necessary to control the fluctuations in the SEIRUC model parameters to make $\mathcal{R}_0<1.$ The sensitivity index of a variable to a parameter is the ratio of the relative change in the variable to the relative change in the parameter, which can be estimated from $S[h]=\frac{h}{\mathcal{R}_0}\times\frac{\partial\mathcal{R}_0}{\partial h}.$ The normalised sensitivity indices of the reproduction number with respect to the system parameters are given in the following table
\begin{center}
	\begin{figure}
		\includegraphics[scale=0.6]{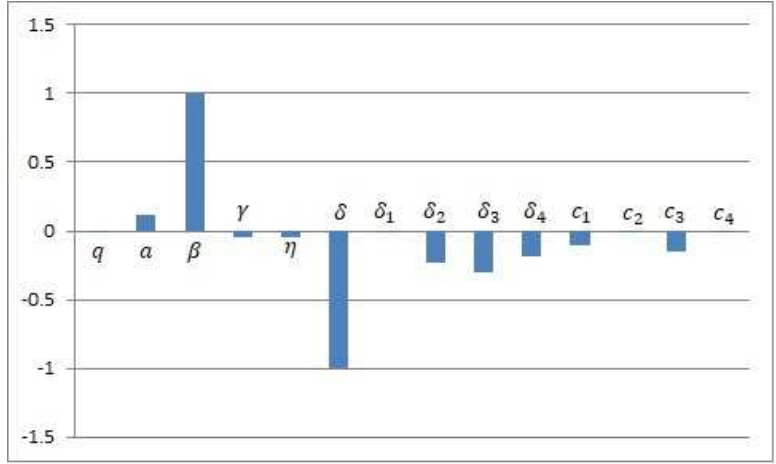}
		\centering
		\caption{Sensitivity indices for the basic reproduction number, $\mathcal{R}_0$ with respect to system parameter values of the SEIRUC model.}
		\label{Fig.Figure1}
	\end{figure}
\end{center}
\begin{center}
	\textbf{Table.1 Sensitivity indices of $\mathcal{R}_0$ with respect to system parameters}\vspace{0.4cm}\label{tab.table1}
\begin{tabular}{|c|l| }	
		\hline
		\textbf{Parameters} & \textbf{Sensitivity index}\\
		\hline
			$S[\delta]$&=$\frac{\delta}{\mathcal{R}_0}\times\frac{\partial\mathcal{R}_0}{\partial\delta}=-1<0$\\
		$S[\beta]$&=$\frac{\beta}{\mathcal{R}_0}\times\frac{\partial\mathcal{R}_0}{\partial\beta}=1>0$\\
		$S[q]$&=$\frac{q}{\mathcal{R}_0}\times\frac{\partial\mathcal{R}_0}{\partial q}=-0.00791883<0$\\
		$S[a]$&=$\frac{a}{\mathcal{R}_0}\times\frac{\partial\mathcal{R}_0}{\partial a}=0.116608>0$\\
		$S[\gamma]$&=$\frac{\gamma}{\mathcal{R}_0}\times\frac{\partial\mathcal{R}_0}{\partial\gamma}=-0.0501895<0$\\
		$S[\eta]$&=$\frac{\eta}{\mathcal{R}_0}\times\frac{\partial\mathcal{R}_0}{\partial\eta}=-0.0487479<0$\\
		$S[\delta_1]$&=$\frac{\delta_1}{\mathcal{R}_0}\times\frac{\partial\mathcal{R}_0}{\partial\delta_1}=-0.0106007<0$\\
		$S[\delta_2]$&=$\frac{\delta}{\mathcal{R}_0}\times\frac{\partial\mathcal{R}_0}{\partial\delta_2}=-0.234848<0$\\
		$S[\delta_3]$&=$\frac{\delta_3}{\mathcal{R}_0}\times\frac{\partial\mathcal{R}_0}{\partial\delta_3}=-0.306767<0$\\
		$S[\delta_4]$&=$\frac{\delta_4}{\mathcal{R}_0}\times\frac{\partial\mathcal{R}_0}{\partial\delta_4}=-0.191107<0$\\
		$S[c_1]$&=$\frac{c_1}{\mathcal{R}_0}\times\frac{\partial\mathcal{R}_0}{\partial c_1}=-0.106007<0$\\
		$S[c_2]$&=$\frac{c_2}{\mathcal{R}_0}\times\frac{\partial\mathcal{R}_0}{\partial c_2}=-0.00757576<0$\\
		$S[c_3]$&$=\frac{c_3}{\mathcal{R}_0}\times\frac{\partial\mathcal{R}_0}{\partial c_3}=-0.148435<0$\\
		$S[c_4]$&$=\frac{c_4}{\mathcal{R}_0}\times\frac{\partial\mathcal{R}_0}{\partial c_4}=-0.0123295<0.$\\
		\hline
		\end{tabular}
\end{center}
\vspace{1.5cm}
From Table 1 and Figure (\ref{Fig.Figure1}), it is obvious that the most sensitive parameter to the basic reproduction $\mathcal{R}_0$ for the SEIRUC model system is the disease transmission rate, $\beta$ and the least sensitive parameter is the natural decay rate of the population, $\delta$. The value of $\mathcal{R}_0$ increases as $\beta$ increases. Thus, $\mathcal{R}_0$ increases proportionally with the increase in transmission rate of infection ($\beta$).
	\section{Numerical Simulation}
	The aim of this study is to determine how the model parameters such as transmission rate, natural decay rate etc,. influence the model framework. Hence, some numerical simulation findings are presented in this section to validate the theoretical results. The model parameter values introduced here are biologically feasible and the data are collected from the literature related to mathematical modelling of various epidemic disease transmission dynamics \cite{FTK3RF10}. We used ode23 MATLAB tool to perform the Numerical simulation of the proposed SEIRUC model. The model parameter values are given in Table 2. \\\\
	\small{\begin{tabular}{ l l l }
		&~~~~~~~~~~~~~~~~~~~\textbf{Table.2} Parameters for model (\ref{e1})&\\	
		\hline
		\textbf{Parameters} & \textbf{Description} & \textbf{Values(unit)}\\
		\hline
		$\Lambda=N_0\times\delta$&Net inflow of Susceptible population&-\\
		$\beta$&disease transmission rate&variable\\
		$\delta$&Natural decay rate of Susceptible population&0.062 day$^{-1}$\\
		$a$&rate at which $E(t)$ becomes $I(t)$&$\frac{1}{12}$day$^{-1}$\\
		$\delta_1$&Natural decay rate of the population $E(t)$&0.001day$^{-1}$\\
		$\delta_2$&Natural decay rate of the population $I(t)$&0.062day$^{-1}$\\
		$\delta_3$&Natural decay rate of the population $R(t)$&0.062day$^{-1}$\\
		$\delta_4$&Natural decay rate of the population $U(t)$&0.062day$^{-1}$\\
		$\delta_5$&Population recovered from COVID-19 but died because of other diseases/reasons&0.061day$^{-1}$\\
		$\frac{1}{\gamma}$&Period of time at which the population $I(t)$ remain asymptomatic&5day$^{-1}$\\
		$q$&Fraction of $I(t)$ that become $R(t)$ &0.1day$^{-1}$\\
		$\eta$&rate at which the population $U(t)$ becomes $R(t)$&$\frac{1}{7}$day$^{-1}$\\
		$c1$&Recovery rate of the population $E(t)$&0.01day$^{-1}$\\
		$c2$&Recovery rate of the population $I(t)$&0.002day$^{-1}$\\
		$c3$&Recovery rate of the population $R(t)$&0.03day$^{-1}$\\	
		$c4$&Recovery rate of the population $U(t)$&0.004day$^{-1}$\\
		$N_0$&Total Population at time $t_0$&1,352,642,280\\
		\hline
	\end{tabular}\label{tab.table2}\\
\\
\indent Figure (\ref{Fig.Figure2}) represents the region plot for the basic reproduction number $\mathcal{R}_0$ with respect to the parameters $\beta$ and $q$ for the region in which $\mathcal{R}_0<1.$ From Figure (\ref{Fig.Figure2}), it can be seen that the basic reproduction number $\mathcal{R}_0$ is less than 1 for the values of disease transmission rate $\beta$ ranging approximately from 0 to 0.1 and $\mathcal{R}_0$ is greater than 1 for the values of $\beta$ greater than 0.1. This  shows that, for higher transmission rates, the disease persists in the population. Hence, we chose the values for $\beta$ between 0 to 0.1 for the endemic-free steady state and greater than 0.1 for the endemic steady state while performing the numerical simulations. \\
\indent For the endemic-free steady state, we consider the parameter values given in Table.1 along with the disease transmission rate $\beta=0.07$. We obtain the endemic-free equilibrium point at $(1.3526\times10^{9},0,0,0,0,0)$. We find the value of $\mathcal{R}_0$ as 0.8005 and the eigenvalues of $J(\mathcal{E}_0)$ as -0.062, -0.061, -0.0094, -0.2736, -0.1881+0.041i and -0.1881-0.041i. It is observed that the given set of model parameters satisfy the local stability conditions for the computed equilibrium points and the endemic-free steady state is locally asymptotically stable for $\mathcal{R}_0<1$. To validate this graphically, the solution trajectories are presented in Figure (\ref{Fig.Figure4}). \\
\indent Figure (\ref{Fig.Figure4}) depicts the local stability of the disease-free equilibrium. The figure shows that the solution trajectories of the SEIRUC model converge to the disease-free equilibrium. Hence, the disease cannot invade the population stating that the disease-free equilibrium is locally asymptotically stable.\\
\begin{figure}
	\includegraphics[width=150mm,height=150mm]{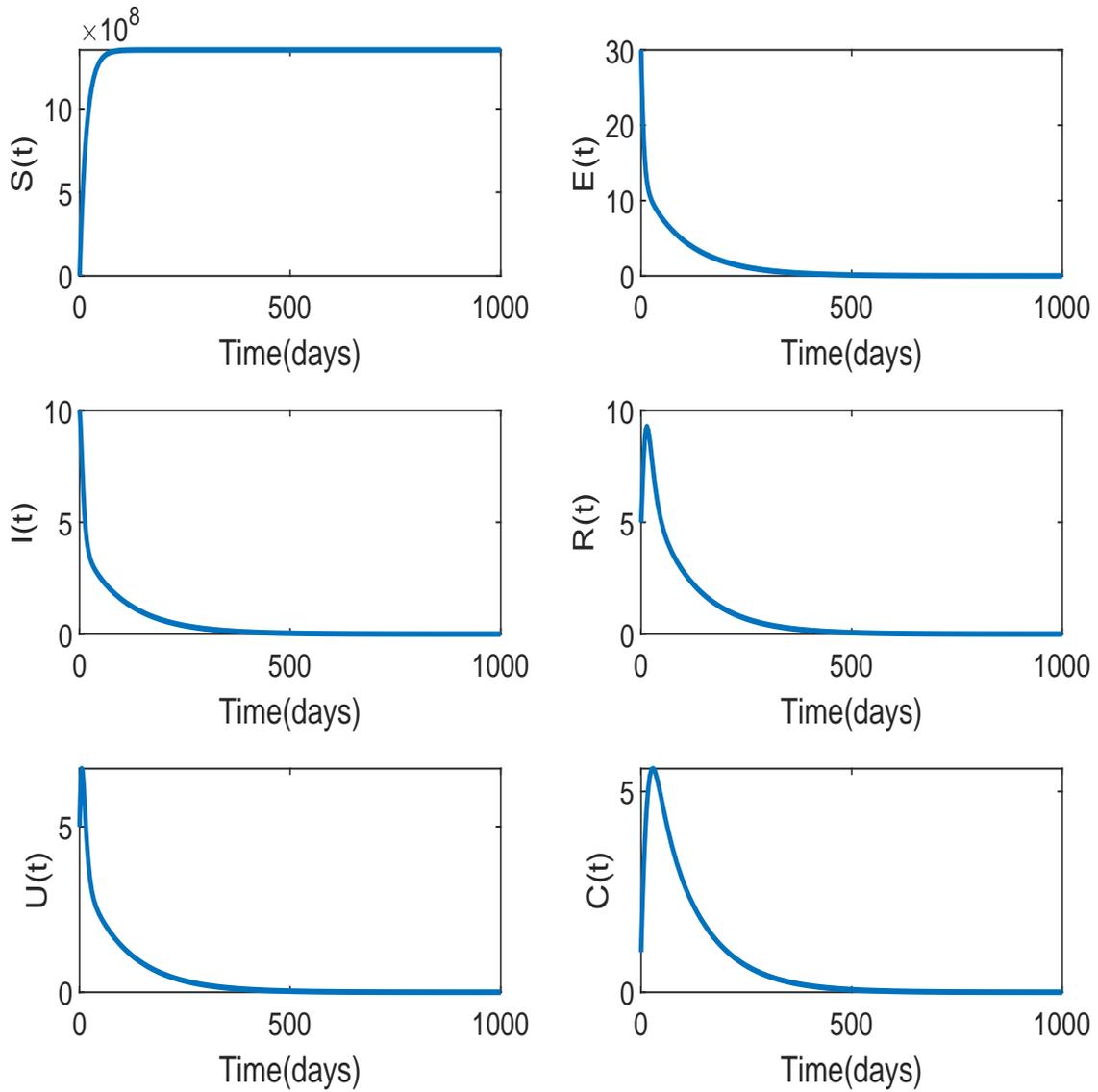}
	\centering
	\caption{Solution trajectories of the endemic-free steady state for the SEIRUC model with $\beta=0.5$ and $\mathcal{R}_0=0.8005<1.$}
	\label{Fig.Figure4}
\end{figure}
\begin{figure}
	\includegraphics[width=150mm,height=150mm]{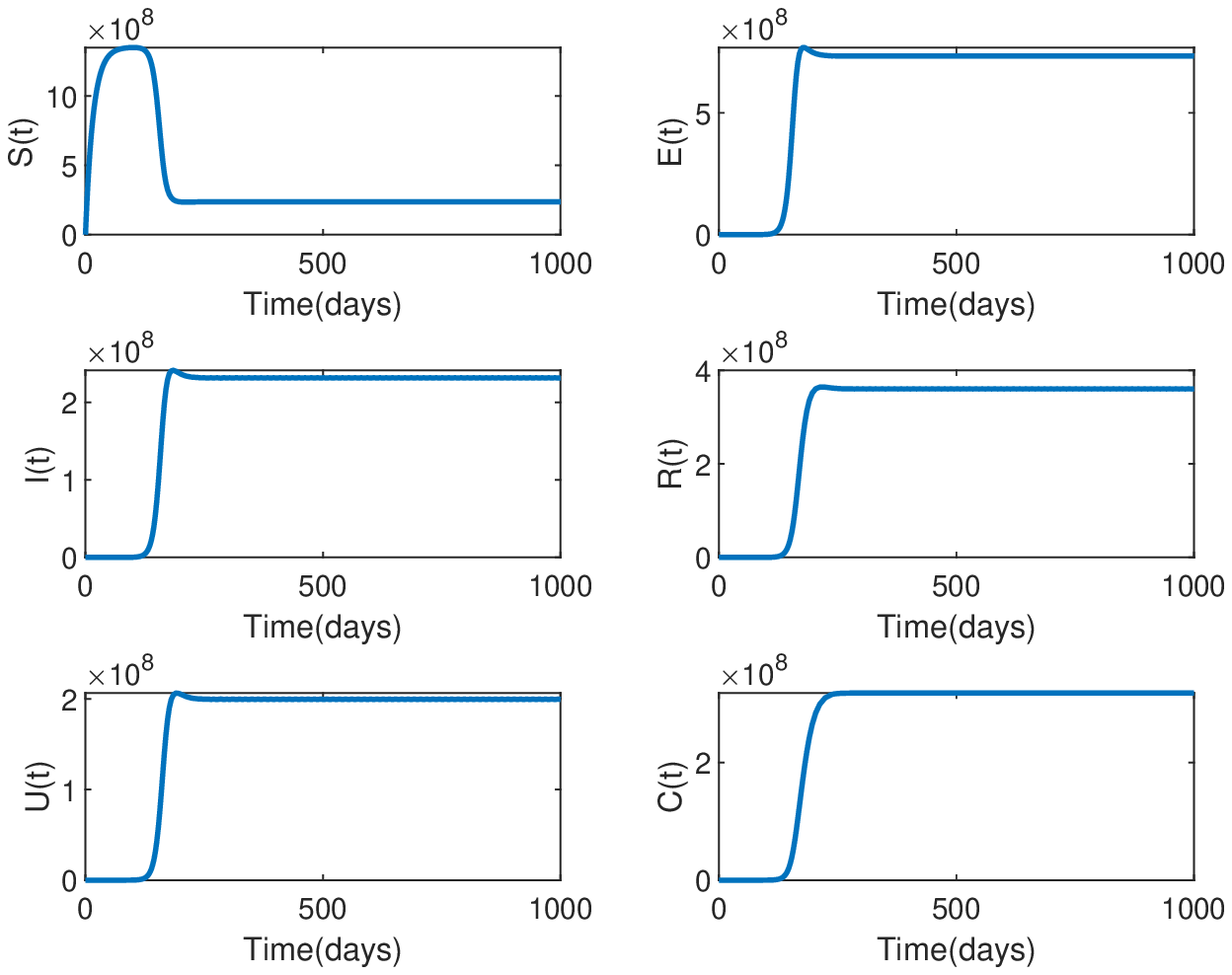}
	\centering
	\caption{Solution trajectories for the SEIRUC model satisfyig stability conditions for the endemic equilibrium point with $\beta=0.5$ and $\mathcal{R}_0=5.7177>1.$}
	\label{Fig.Figure5}
\end{figure}
For the endemic steady state, we consider the model parameter values given in Table.2 along with disease transmission rate $\beta=0.5$, where we obtain the value of $\mathcal{R}_0$ as 5.7177. We obtain the endemic equilibrium point at $(2.36569,7.33532,2.31544,3.6020,1.99553,3.18076)\times 10^8$. The eigenvalues of the corresponding Jacobian matrix $J(\mathcal{E}_*)$ are -0.0610, -0.0749+0.0253i, -0.0749-0.0253i, -0.2408+0.0231i, -0.2408-0.0231i and -0.3822. Hence, all the eigenvalues of $J(\mathcal{E}_*)$ are found to be having negative real parts establishing stability of the proposed system at the endemic equilibrium point. The solution trajectories satisfying the local stability condition for  endemic steady state can be seen in Figure (\ref{Fig.Figure5})\\
\indent Figure (\ref{Fig.Figure5}) shows that the susceptible population decreases with increase in time and the infectious compartments increases with increase in time. The solution trajectories of system (\ref{e1}) converge to the endemic equilibrium point showing the persistence of the disease in the population, which indicates the real situation that the world is still fighting against this deadly virus with local stability.\\
\indent In order to better understand the transmission dynamics of our SEIRUC model, In Figure (\ref{Fig.Figure3}) we present a 3-dimensional plot for the basic reproduction number, $\mathcal{R}_0$ with respect to the rate of transmission of disease, $\beta$ and the rate at which the asymptomatic infectious individuals turns into reported symptomatic individuals $(q)$.\\
\newpage
\section{Conclusion}
We have proposed a new mathematical model by considering the new class called convalesce class along with asymptomatic infectious and symptomatic infectious class for the transmission dynamics of COVID-19. By exploiting Routh-Hurwitz criteria for higher order polynomials, we established the sufficient conditions for the local stability of the disease-free and endemic equilibrium. Using the concept of next generation matrices, the threshold quantity, $\mathcal{R}_0$ has been computed. The established qualitative behaviour of the proposed model has been verified by the numerical simulations.\\
\indent In future work, we aim to study the qualitative behaviour and numerical aspects of the proposed SEIRUC model to understand the transmission dynamics of COVID-19 under different fractional order derivatives.
	\section*{Acknowledgement}
	This research work is supported by National Board for Higher Mathematics, Department of Atomic Energy, Mumbai, India under the grant no. 02011/8/2020/NBHM/R$\&$D-II/8071.
	
\end{document}